\documentclass[letterpaper,10pt]{article} 

\usepackage{opticameet3} 

\newcommand\authormark[1]{\textsuperscript{#1}}

\usepackage{amsmath,amssymb}
\usepackage[colorlinks=true,bookmarks=false,citecolor=blue,urlcolor=blue]{hyperref} 

\usepackage{xcolor}

\begin{document}

\title{Simulating Circuit Layout for Distributed Quantum Computing}


\vspace{-15pt}

\author{Sen Zhang\authormark{1}, Yipei Liu\authormark{1}, Brian Mark\authormark{1}, Weiwen Jiang\authormark{1}, Zebo Yang\authormark{2}, Lei Yang\authormark{1}}

\address{\authormark{1}George Mason University, Fairfax, VA 22030, USA \ \ \ \ \authormark{2}Florida Atlantic University, Boca Raton, FL 33431, USA}






\vspace{-10pt}

\begin{abstract}
The proposed framework represents the first tool to compile a quantum circuit across photonic-connected distributed quantum processors. Its design follows a divide-and-conquer paradigm for circuit partitioning, transpilation, and assembly, producing simulatable and implementable circuit layouts.
\vspace{-15pt}
\end{abstract}

\section{Introduction}
Current Quantum Processing Units (QPUs), such as those from IBM, Regitti, IonQ, Quantinuum, QuEra, and Infleqtion, lack the capacity to support the scale and complexity of quantum circuits (in particular, for the fault-tolerant quantum computing circuits) required for practical applications, challenging scalability and reliability \cite{cuomo2020towards}. Monolithically scaling a QPU is extremely difficult due to fabrication limits and errors. Distributed Quantum Computing (DQC) offers a promising solution by interconnecting multiple QPUs through quantum photonic links and optical switching networks to collaboratively solve a single problem at scale \cite{8910635}. The heterogeneity of distributed QPUs, across superconducting, trapped-ion, and photonic platforms, together with the complexity of optical links and quantum repeaters, introduces significant challenges for DQC system design and verification \cite{CALEFFI2024110672}. 
Given the rapid progress in DQC, it is essential to initiate research efforts during the interim period while the hardware ecosystem continues to mature.
However, the field lacks a comprehensive validation platform to support the development and testing of DQC strategies, as well as the benchmarking quantum algorithms in distributed settings. 

The cornerstone for quantum validation platforms or simulator is the quantum circuit layout \cite{tan2020optimal}, which specifies how logical qubits and gates in a circuit are mapped and arranged on given quantum hardware.
Unlike targeting a single QPU, the layout across distributed QPUs faces a set of new needs and challenges.
\textit{First}, the distributed quantum circuit layout requires a circuit-level modeling for the communication (more specifically, the two-qubit gates) between a pair of QPUs.
\textit{Second}, it requires QPU backend-specified compilation, as heterogeneous QPUs have different connectivity topologies and basis gates for implementation. 
\textit{Moreover}, due to inter-QPU communication, the compilations among QPUs require synchronization to ensure the correctness of the function.
All the above challenges cannot be simply addressed by revising an existing quantum circuit layout for a single QPU.


To address this pressing issue, we introduce our framework, the first automated layout compilation tool that accounts for the heterogeneity of QPUs in DQC, providing an end-to-end automated compilation to generate the layout of a given quantum circuit across distributed QPUs, which can be both simulatable and implementable.
The design of it follows \textbf{the divide-and-conquer paradigm}.
Specifically, in the divide step, it begins by partitioning a quantum circuit into subcircuits, which are then mapped to a predefined QPU network topology. 
Then, in the conquer step, each subcircuit is independently compiled according to the hardware specifications of its assigned QPU, such as qubit connectivity and supported gate sets.
Finally, in the combine step, following the compiled subcircuits, an automated placement algorithm will be developed to assemble these subcircuits with additional remote quantum gates to generate a fully instantiated circuit, where 
Our framework seamlessly manages entanglement distribution and remote gate execution through quantum networking and TeleGate \cite{PhysRevA.62.052317,ferrari2021compiler}. 
The generated quantum circuit layout can enable users to simulate a quantum circuit across multiple heterogeneous backends to assess circuit performance and correctness under realistic DQC conditions; meanwhile, it is also implementable if Einstein–Podolsky–Rosen (EPR) pairs can be established on physical quantum computers.


\begin{figure}[t]
  \centering
  \includegraphics[width=1\textwidth]{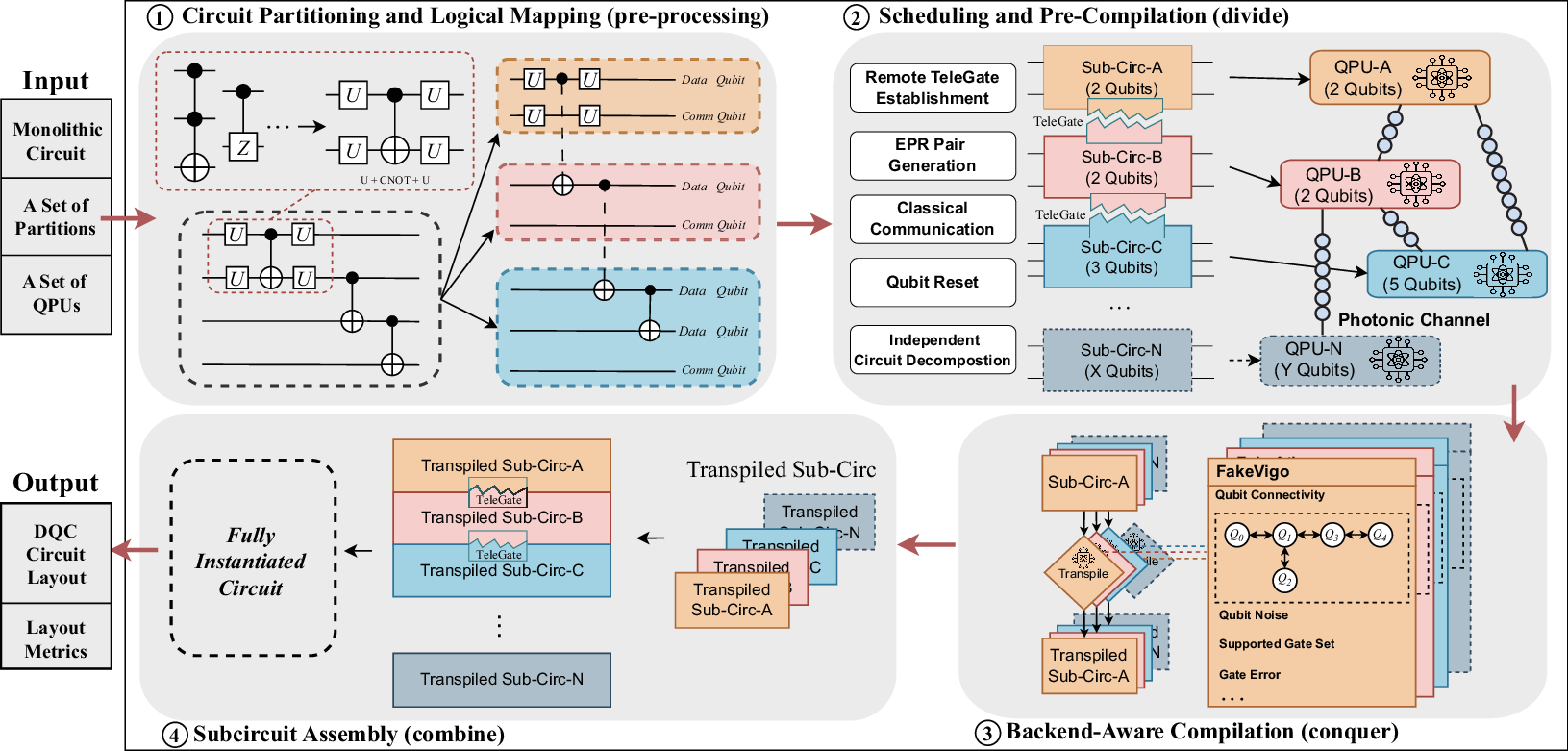}
\caption{End-to-End Design Flow and Architecture of Our Framework.}
\label{fig:workflow}
\end{figure}

\begin{figure}[b!]
  \centering
  \includegraphics[width=1\textwidth]{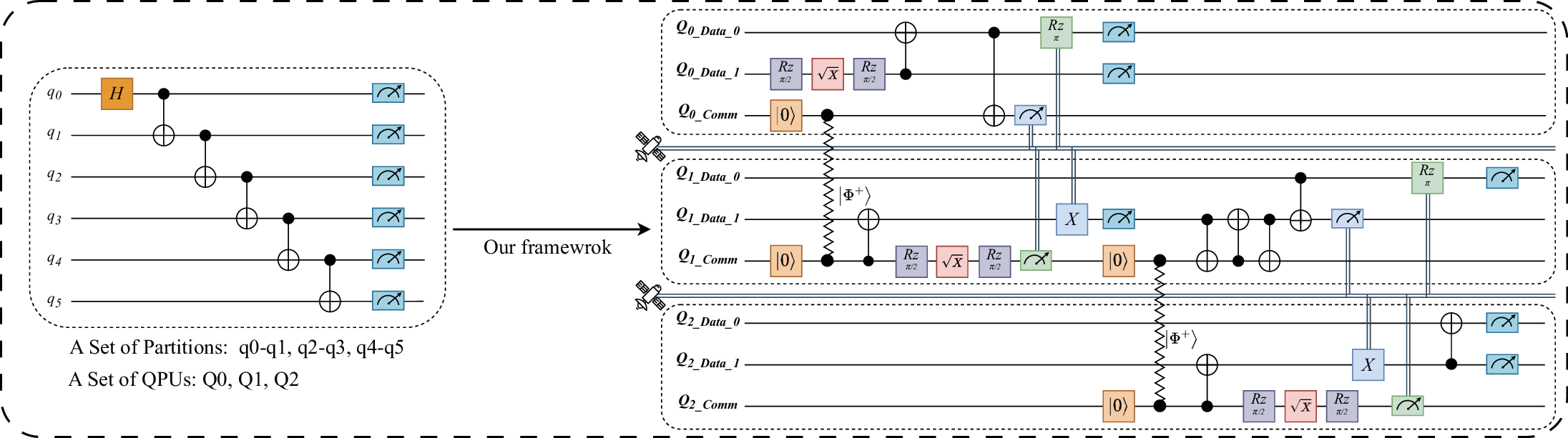}
\caption{Example of Distributed Layout Compilation for GHZ6 Circuit using Our Framework.}
\label{fig:layoutCompilation}
\end{figure}

\section{Proposed Framework}

Figure~\ref{fig:workflow} illustrates the end-to-end architecture of our framework.
Given a set of QPUs connected via photonic channels, a monolithic quantum circuit, and a set of user-specified qubits partitions, as shown in the Input in the figure, our framework can automatically transform the quantum circuit into a fully simulatable and implementable distributed layout, as shown in the Output of the figure.
In addition, a set of associated metrics, such as different types of qubits and circuit depth, will be reported in the output.
As introduced, the workflow of our framework follows the divide-and-conquer paradigm, including four main stages, detailed below.

\textbf{\ding{172} Circuit Partitioning and Logical Mapping (pre-processing)}:
The given circuits can contain different types of multi-qubit gates, such as Toffoli and CZ gates. However, the TeleGate execution can only support a limited set of gates, such as a remote CNOT gate.
Therefore, the pre-processing step is to use the partitions given by the user to convert multi-qubit gates across partitions to supported remote gates, as shown in the red dashed box in Figure \ref{fig:workflow} \ding{172}.
Based on the remote gate conversion, this step further parses each partition sub-circuit into a logical group with additional communication qubits.
These groupings define the logical boundaries for distributed execution.

\textbf{\ding{173} Scheduling and Pre-Compilation (divide)}: 
With the logical groups, this divide step aims to isolate the subcircuits onto different QPUs, such that they can be compiled independently on QPUs in the next step.
This is motivated by two facts: (1) existing compilation tools do not support remote gates; and (2)  each QPU may have different capabilities (e.g., qubits, connectivity), which prefer independent compilation.
To achieve this, we first assign subcircuits to specific QPUs with consideration of resource limitations. 
Then, 
remote TeleGates are explicitly scheduled to bridge operations between QPUs, including the establishment of EPR pairs between QPUs, coordinating classical communication, and resetting qubits post-execution. 
Finally, to isolate different sub-circuits, we replace remote gates with placeholders, which ensures they will not be compiled with other local gates.


\textbf{\ding{174} Backend-Aware Compilation (conquer)}: This conquer step is to perform backend-aware transpilation of a subcircuit based on its assigned QPU's characteristics, including native gate sets and qubit topology. It ensures that the compiled circuit segments are executable on heterogeneous backends, including real devices (e.g., IBM, IonQ, or others) or simulated devices (e.g., FakeVigo). 

\textbf{\ding{175} Subcircuit Assembly (combine)}: This is the key step to establish
a fully instantiated distributed circuit. 
This assembly process leverages the placeholders in step \ding{173} that
represent cross-QPU interactions, which are now resolved back into remote gate operations, i.e., TeleGate. These remote gates restore the logical dependencies across subcircuits, reconnecting the independently compiled subcircuits into a coherent global circuit.

Figure~\ref{fig:layoutCompilation} presents an example of compilation process applied to a 6-qubit GHZ state generation circuit. It demonstrates how a monolithic quantum circuit is transformed into a distributed execution layout across multiple QPUs, each with its own local qubit resources and communication channels. The left panel shows the inputs, including a standard 6-qubit GHZ circuit, the user-specified partition and a set of QPUs.
The right panel shows the resulting distributed quantum circuit layout over three QPUs labeled $Q_0$, $Q_1$, and $Q_2$, each with distinct data and communication qubits. Each QPU handles a partitioned subcircuit, with computation localized to available data qubits. Remote CNOTs based on Bell pairs ($\left|\Phi^+\right\rangle$ and helical wires) bridge operations across QPUs. Native gates (e.g., $R_z$, $\sqrt{X}$) are generated where appropriate.


\section{Simulation and Performance Results}

\begin{table*}[t]
\centering
\scriptsize
\caption{Performance Evaluation of Distributed Quantum Circuit Benchmarks.}
\label{tab:ComparisonOfQuantumCircuitSets}
\renewcommand{\arraystretch}{1.08}
\setlength{\tabcolsep}{0.92pt}
\begin{tabular}{|c|ccc|ccc|c|c|c|ccc|c|c|c|c|c|c|}
\hline
\multirow{3}{*}{\textbf{Benchmark}} 
& \multicolumn{6}{c|}{\textbf{Input}} 
& \multicolumn{8}{c|}{\textbf{Circuit Metrics}} 
& \multicolumn{4}{c|}{\textbf{Fidelity}} \\
\cline{2-19}
& \multicolumn{3}{c|}{Partition} 
& \multicolumn{3}{c|}{Assignment} 
& \multirow{2}{*}{\#QData} 
& \multirow{2}{*}{\#QComm} 
& \multirow{2}{*}{\#QTotal} 
& \multicolumn{3}{c|}{Sub-Circ Depth} 
& Layout
& Gate
& \multirow{2}{*}{State$^{*}$} 
& \multirow{2}{*}{IProb} 
& \multirow{2}{*}{EProb} 
& \multirow{2}{*}{Error Rate} \\
\cline{2-4} \cline{5-7} \cline{11-13}
& P0 & P1 & P2 
& P0 & P1 & P2 
&  &  &  
& Min & Max & Avg 
& Depth & Count &  &  &  &  \\
\hline
GHZ-6 & q0-q1 & q2-q3 & q4-q5 & Q0 & Q1 & Q2 & 6 & 4 & 14 & 5 & 10 & 7.33 & 12 & 30 & $|0\rangle^{\otimes 6}$ & 0.5 & 0.50144 & $10^{-6}$ \\
GHZ-6 & q0-q1 & q2-q3 & q4-q5 & Q0 & Q2 & Q1 & 6 & 4 & 14 & 5 & 10 & 7.33 & 12 & 30 & $|0\rangle^{\otimes 6}$ & 0.5 & 0.49861 & $10^{-6}$ \\
GHZ-6 & q0 & q1-q2 & q3-q5 & Q0 & Q2 & Q1 & 6 & 4 & 14 & 5 & 10 & 7.67 & 13 & 30 & $|0\rangle^{\otimes 6}$ & 0.5 & 0.50106 & $10^{-6}$ \\
GHZ-12 & q0-q3 & q4-q7 & q8-q11 & Q0 & Q1 & Q2 & 12 & 4 & 16 & 7 & 11 & 9 & 17 & 42 & $|0\rangle^{\otimes 12}$ & 0.5 & 0.49863 & $10^{-6}$ \\
BitCode-3 & q0-q2 & q3-q4 & - & Q0 & Q1 & - & 5 & 4 & 9 & 9 & 15 & 12 & 14 & 40 & $|00100\rangle$ & 1 & 1 & 0 \\
BitCode-3 & q0 & q1-q4 & - & Q0 & Q1 & - & 5 & 4 & 9 & 9 & 17 & 13 & 16 & 40 & $|00100\rangle$ & 1 & 1 & 0 \\
TFIM Hamiltonian & q0 & q1-q2 & - & Q0 & Q1 & - & 3 & 12 & 15 & 28 & 43 & 35.5 & 43 & 102 & $|1\rangle^{\otimes 3}$ & 0.6 & 0.60458 & $10^{-5}$ \\
TFIM Hamiltonian & q0-q1 & q2 & - & Q0 & Q1 & - & 3 & 12 & 15 & 34 & 37 & 35 & 43 & 102 & $|1\rangle^{\otimes 3}$ & 0.6 & 0.60413 & $10^{-5}$ \\
Qaoa-4 & q0-q1 & q2-q3 & - & Q0 & Q1 & - & 4 & 16 & 20 & 36 & 44 & 40 & 45 & 102 & $|0001\rangle$ & 0.093612 & 0.094750 & $10^{-5}$ \\
Qaoa-6 & q0-q1 & q2-q3 & q4-q5 & Q0 & Q1 & Q2 & 6 & 48 & 54 & 67 & 83 & 75.33 & 122 & 279 & $|001101\rangle$ & 0.048330 & 0.048520 & $10^{-4}$ \\
Qaoa-8 & q0-q2 & q3-q5 & q6-q7 & Q0 & Q1 & Q2 & 6 & 84 & 90 & 123 & 140 & 130 & 205 & 486 & $|01110101\rangle$ & 0.028230 & 0.028270 & $10^{-3}$ \\
Qaoa-10 & q0-q3 & q4-q6 & q7-q9 & Q0 & Q1 & Q2 & 10 & 132 & 142 & 194 & 213 & 201.33 & 319 & 759 & $|0000001011\rangle$ & 0.007284 &  0.006950 & $10^{-3}$ \\
\hline
\end{tabular}
\vspace{2pt}
\noindent
\parbox{\linewidth}{%
\footnotesize ``*'': Due to space limitations, for each benchmark, we only present the result of the state with the highest probability in output distribution.}
\end{table*}

To validate the resultant layout, we implement our framework using IBM Qiskit \cite{qiskit2024}, and the layouts are simulated using IBM Aer simulator .
Table~\ref{tab:ComparisonOfQuantumCircuitSets} shows the results of different distributed quantum circuit benchmarks executed using our framework. We evaluate four representative quantum algorithms: the GHZ circuit for state preparation, BitCode circuit for syndrome measurement in bit-flip quantum error correction code, TFIM (Transverse Field Ising Models) circuit for Hamiltonian simulation, and the QAOA for optimization problems.
The circuits of BitCode, TFIM, and QAOA are obtained from the SupermarQ benchmark suite \cite{tomesh2022supermarq}.

In the table, the \textit{Input} columns list the user-defined partitions and backend assignments. Here, Q0, Q1, and Q2 correspond to the quantum backends FakeVigoV2, FakeAthensV2, and FakeLagosV2, respectively.
The \textit{Circuit Metrics} columns summarize the key configurations of the circuit layouts. 
\textit{\#QComm} denotes the number of communication qubits, which corresponds in practice to EPR pairs established through optical links.
The \textit{Sub-Circ Depth} reports the depth of each subcircuit on the assigned QPU without remote gates, while the \textit{Layout Depth} is for the resultant layout.
\textit{Gate Count} lists the number of quantum gates within the layout. 

From the table, we observe that all circuit metrics grow proportionally, as the complexity of the quantum algorithm and the number of qubits increase. Notably, the total number of gates in the simulatable layouts generally exceeds that of the subcircuits, highlighting the additional overhead of distributed quantum computation. 
For identical benchmarks, variations in partitioning strategies result in significant differences between the Sub-Circ Depth and the final Layout Depth. This behavior, observed in the BitCode, TFIM Hamiltonian, and GHZ circuits, underscores the critical role of circuit partitioning in the overall performance of distributed quantum computation.


The \textit{Fidelity} column reports the accuracy of the circuits generated by our framework.
The results show that after heterogeneous compilation, the layout fidelity remains nearly identical to the ideal state, validating the correctness of the our workflow.
Here, \textit{IProb} denotes the ideal probability, while \textit{EProb} shows the experimentally obtained probability. The error rate is defined as $1 - f$, where $f$ is the Hellinger Fidelity, which quantifies the similarity between two probability distributions.
For QAOA, the error rate increases with the number of qubits because larger circuits take a longer time to simulate with a larger number of measurement shots. With fewer shots, the output probabilities become less accurate, leading to higher measured error rates. 


\section{Conclusion}

Our framework achieves near-perfect results in the noiseless setting, demonstrating the effectiveness of distributed quantum execution across heterogeneous QPUs. As the first simulator designed for distributed quantum circuit layouts, it serves as a valuable development framework for DQC researchers to benchmark their own solutions and establishes a solid foundation for future research in this field.


\bibliographystyle{opticajnl}
\bibliography{sample}

\end{document}